\documentclass[preprint,review,11pt]{elsarticle}

\makeatletter
\def\ps@pprintTitle{%
 \let\@oddhead\@empty
 \let\@evenhead\@empty
 \def\@oddfoot{}%
 \let\@evenfoot\@oddfoot}
\makeatother

\usepackage{lineno}\modulolinenumbers[5]
\usepackage{amsmath}
\usepackage{xcolor, soul}
\usepackage{graphicx}
\usepackage{tikz}
\usetikzlibrary{spy}
\usepackage{pgfplots}\pgfplotsset{compat=1.16,filter discard warning=false,unbounded coords=jump}\usepgfplotslibrary{groupplots}
\usepackage{booktabs}
\usepackage{amsthm}
\usepackage[frozencache=true,cachedir=.]{minted}

\newcommand{\appropto}{\mathrel{\vcenter{
  \offinterlineskip\halign{\hfil$##$\cr
    \propto\cr\noalign{\kern2pt}\sim\cr\noalign{\kern-2pt}}}}}

\journal{Journal of Computer and System Sciences}

\begin{document}

\begin{frontmatter}

\title{About Weighted Random Sampling in Preferential Attachment Models}

\author[]{Giorgos~Stamatelatos\corref{mycorrespondingauthor}}
\cortext[mycorrespondingauthor]{Corresponding author}
\ead{gstamat@ee.duth.gr}
\author[]{Pavlos~S.~Efraimidis}
\ead{pefraimi@ee.duth.gr}
\address{Dept. of Electrical and Computer Engineering, Democritus University of Thrace, Kimmeria, Xanthi 67100, Greece}

\begin{abstract}
The Barabási--Albert model is a popular scheme for creating scale-free graphs but has been previously shown to have ambiguities in its definition. In this paper we discuss a new ambiguity in the definition of the BA model by identifying the tight relation between the preferential attachment process and unequal probability random sampling. While the probability that each individual vertex is selected is set to be proportional to their degree, the model does not specify the joint probabilities that any tuple of $m$ vertices is selected together for $m>1$. We demonstrate the consequences using analytical, experimental, and empirical analyses and propose a concise definition of the model that addresses this ambiguity. Using the connection with unequal probability random sampling, we also highlight a confusion about the process via which nodes are selected on each time step, for which -- despite being implicitly indicated in the original paper -- current literature appears fragmented.
\end{abstract}


\end{frontmatter}

\section{Introduction}
\label{sec:introduction}

\paragraph{The Barabási--Albert model}

The Barabási--Albert model (BA model) \citep{barabasi1999emergence} is a very popular model for generating undirected scale-free networks, i.e. graphs with power-law degree distribution. This property is achieved via a growing scheme with a preferential attachment mechanism, in which each newborn node gives links to existing nodes and the more connected a node is the more likely it is to receive a new link. More formally, at each time step $t$ the new node $t$ that is born gains $m$ different edges to existing nodes. According to the original paper, for the selection of the $m$ different nodes we (quoted with adjusted notation)
\begin{quote}
randomly select a vertex $i$ and connect it with probability $m \cdot d_i(t) / \sum_{j=1}^t d_j(t)$ to vertex $t$ in the system,
\end{quote}
where $d_i(t)$ is the degree of vertex $i$ at time $t$.

The BA model is also described in textbooks, for example \citet{jackson2010social} offers a slight variation of the original definition:
\begin{quote}
the probability that an existing node $i$ gets a link from the new born node at time $t$ is $m$ times $i$'s degree relative to the overall degree of all existing nodes at time $t$, or $m \cdot d_i(t) / \sum_{j=1}^t d_j(t)$.
\end{quote}
The model is also widely used in research works with tens of thousands of citations and is implemented in popular network libraries, such as NetworkX (Python), jGraphT (Java) and iGraph (C++).

Following the original definition of the model, the process of selecting $m$ nodes from the population based on their degrees is referred to as \textit{weighted random sampling}~\cite{efraimidis2006weighted} or \textit{unequal probability random sampling}~\cite{tille2006sampling} or \textit{varying probability random sampling}~\cite{arnab_survey_2017}. More formally, it consists in selecting $m$ items from a population of $t$ items where the probability of selecting any two items may not necessarily be equal or in which all possible samples need not have the same selection probabilities. Typically, in the preferential attachment models, these probabilities are functions of the elements' sizes, in the case of the BA model the elements' degrees. To the best of our knowledge, this is the first time that the growing preferential attachment model is studied in relation to the unequal probability random sampling schemes and, as a result, the first time that preferential attachment is identified as an application of unequal probability random sampling. This subject is further analyzed in Section~\ref{sec:random-sampling}. This relation motivates us to study the BA model in this paper in a perspective that is absent in current literature but is fundamental in a random sampling context: the \textit{inclusion probabilities} of elements (vertices) to be included in the final $m$ vertices of the sample. More specifically, we distinguish these into the inclusion probabilities of individual elements (first order) and the joint inclusion probabilities of groups of elements (higher order) to be included in the final sample of $m$ vertices.

\paragraph{First order inclusion probabilities}

The first order inclusion probability of an item is defined as the number of samples that contain that element over the number of all possible samples; this is equivalent to the probability of that item to exist in the final selected sample. According to the proof given in the original paper~\cite{barabasi1999emergence} as well as the follow up work~\cite{albert2002statistical}, it is indirectly stated that the sampling model corresponds to the one with inclusion probabilities strictly proportional to size (str$\pi$ps). Despite this, we discover that most research studies about preferential attachment and the BA model employ sampling models that are different than the first order str$\pi$ps model that, for example, result in multigraphs or other models that are only approximations of str$\pi$ps. Most notably, the majority of work around the BA model interprets the inclusion probabilities of an item as the \textit{selection probabilities} of that item to be selected at any of the indirect $m$ individual and independent selections. This sampling scheme is different than the strict model defined in the original work of Barabási and Albert and is only an approximation of it for finite $n$. The two models are equivalent only for $m=1$. We also discover that open source implementations of the BA model also suffer from this issue as most implement the incorrect selection probabilities scheme. These findings are more thoroughly presented in Section~\ref{sec:related-work}.

Following this confusion that exists in the literature regarding the first order inclusion probabilities of vertices, we perform computer experiments in an attempt to quantify differences among the sampling designs. We identify multiple perspectives via which the designs differ and present the findings in Section~\ref{sec:first-order}. In that section we selected 3 random sampling designs with different first order inclusion probabilities and compare them in attempt to show the practical implication of this distinction. The examined cases include the designs of the inclusion and selection probabilities mentioned before.

\paragraph{Higher order inclusion probabilities}

In a random sampling scheme perspective, the higher order probabilities refer to the joint inclusion probabilities of tuples of items. In the context of the BA model the $k$-order inclusion probability of an unordered tuple of vertices corresponds to the probability of all $k$ vertices in that tuple to gain an edge from the newborn node at the same time step. Despite the original definition of the BA model having specified (even indirectly) the first order inclusion probabilities, the specification of the higher order probabilities is not specified. Furthermore the higher order probabilities cannot be deduced by the first order ones and there exist many higher order probability combinations that can yield the same first order ones. This constitutes a new ambiguity in the definition of the BA model and the preferential attachment model in general, the discovery of which and the assessment of its impact has not been discussed in the literature before. In Section~\ref{sec:higher-order}, we analytically show that two sampling methods with equivalent first order probabilities but different higher order probabilities can result in at least one asymptotic difference among the respective growing graph generators. Finally, in Section~\ref{sec:concise-definition} we provide a concise definition of the BA model that addresses this ambiguity by considering the higher order probabilities.

\paragraph{Contribution summary}

In this paper, we investigate the BA model in the perspective of the preferential attachment sampling inclusion probabilities. The analysis of the inclusion probabilities is fundamental to any random sampling design and, consequently the BA model, which is identified here as an application of weighted random sampling. In particular, we study the cases of both first order and higher order probabilities and show that a) most research and open source implementations do not employ exactly the probability model specified in the original BA paper and b) there is an important ambiguity regarding the higher order inclusion probabilities in the definition of the model. With this work, we hope to motivate the further integration of weighted random sampling and preferential attachment in future studies.

Our main contribution can be summarized as follows:
\begin{enumerate}
    \item We associate two scientific problems: unequal probability random sampling and growing preferential attachment models as an application of the former.
    \item Using this association, we describe the confusion about the first order inclusion probabilities of the preferential attachment process and demonstrate its impact.
    \item We identify the ambiguity on the definition of the BA model regarding the higher order inclusion probabilities and establish a strict definition that considers unequal probability random sampling.
\end{enumerate}

\section{Unequal probability random sampling}
\label{sec:random-sampling}

The tight connection between the BA model and weighted random sampling emerges via the original definition of the BA model that was given in~\cite{barabasi1999emergence}, where the authors state the famous ``probability proportional to degree'' principle to create scale-free graphs, i.e. graphs with power law degree distribution. However, the nature of this definition, which is based on the phrase ``select node $i$ with probability $p_i$'', may be interpreted in at least 3 different ways:
\begin{enumerate}
    \item Each individual node $i$ will obtain a link with an \textit{independent} probability $p_i$.
    \item Each node $i$ has a \textit{selection} probability $p_i$ in a scheme where the $m$ nodes are selected one by one until we have selected $m$ nodes in total.
    \item The \textit{inclusion} probability of $i$ to appear in the collection of $m$ final nodes is $p_i$.
\end{enumerate}
These interpretations are just 3 examples that are not equivalent to each other but all abide by the ``probability proportional to degree'' requirement imposed by this phrase. This subtle characteristic may initially create a confusion regarding the exact process of node selection during each step of the growing process, which can lead to discrepancies in both the model analyses and the respective implementations.

The follow up work by the same authors~\cite{albert2002statistical} indirectly suggested that the proposed model utilizes the inclusion probability design. In particular, this observation is evident via the two proofs regarding the degree distribution of the model in Section VII.B; using a continuum approximation and the master equation approach. Both methods assume that the rate at which a node obtains edges is proportional to its degree after the addition of all $m$ new edges, which -- indirectly but clearly -- refers to the inclusion probability interpretation (3). This realization motivates us to study the BA model in relation to weighted random sampling and more precisely the inclusion probabilities, a concept that is fundamental to any random sampling design. The aforementioned proofs show the power law nature of the degree distribution, which was initially observed by Yule~\cite{doi:10.1098/rstb.1925.0002} and later named the \textit{Yule--Simon distribution}:
\[
  f(k;\rho) = \frac{\rho~\rho!~(k-1)!}{(k+\rho)!}.
\]
Albert and Barabási indirectly showed that for their model $\rho=2$, thus
\[
  f(k) \propto \frac{1}{k(k+1)(k+2)},
\]
which for sufficiently large $k$ (tail of the distribution) is a power law
\[
  f(k) \appropto k^{-3}.
\]

Our notation is established by defining the problem of unequal probability random sampling as selecting $m$ elements from a population of $t$ elements based on their sizes $x_1, x_2, ..., x_t$. The values $x$ are also called parameters or weights and in this context they refer to vertex degrees. In addition, the definitions and abbreviations in the rest of this paper are commonly found in the literature~\cite{hanif1980sampling,rosen1997sampling,tille2006sampling}.

The presence of the size parameters in weighted random sampling designs, their interpretation and the exact mechanism via which they are being utilized is the reason of existence of various designs. The same situation is not relevant in unweighted random sampling as there are no size parameters. In the case of growing preferential attachment, during time $t$ when there are $t$ elements in the system, \textit{exactly} $m$ \textit{discrete} elements will be selected based on their existing sizes (degrees). It is, therefore, evident that this refers to a random sampling without replacement and with fixed sample size. As stated earlier, however, a major difference among the designs is the exact process via which the parameters $x_1, x_2, ..., x_t$ are being utilized. In particular, each design interprets the sizes $x_i$ in a different way which leads to differences in the respective inclusion probabilities. More formally, the value $\pi_i$ is defined as the first order inclusion probability of element $i$ in the sample, which is equal to the sum of the probabilities of appearance of all the samples that contain $i$. While the inclusion probabilities are a function of the sizes, a special case of random sampling design is the str$\pi$ps (inclusion probability strictly proportional to size) for which they are exactly proportional ($\pi_i \propto x_i$) while the Horvitz--Thompson estimator for this case is unbiased.

Several unequal probability sampling designs are given by \citet{hanif1980sampling}, \citet{brewer_sampling_1983}, \citet{tille2006sampling}, \citet{berger2009sampling} and \citet{grafstrom2010unequal}. In this work, we discuss the preferential attachment mechanism in relation to three relevant designs but our arguments are more general and hold for a larger selection of sampling methods:
\begin{enumerate}
    \item The conditional Poisson design \citep{hajek1964asymptotic}.
    \item The draw-by-draw selection \citep{yates1953selection}.
    \item The str$\pi$ps scheme.
\end{enumerate}
These designs all refer to unequal probability sampling without replacement and with constant sample size but the interpretation of the weights is differentiated. Specifically, the conditional Poisson is equivalent to the process of generating Bernoulli samples until the desirable sample size is achieved via the rejection of invalid samples. The Yates-Grundy draw-by-draw design is the process of selecting the individual units with unequal probabilities and without replacement until the sample is of the desired size. Efficient algorithms implementing the draw-by-draw design have been suggested, for example in~\cite{efraimidis2006weighted} for which the equivalence has been proven by \citet{li1994computer}. The cases of the draw-by-draw and str$\pi$ps design are also discussed in \cite{efraimidis2015weighted}.

\begin{table}
    \centering
    \begin{tabular}{cccc}
        \hline
        Design & Incl. prob. of node 1 & Difference & Ratio \\
        \hline
        str$\pi$ps & $\frac{4}{n+1}$ & 0 & 1 \\
        Draw-by-draw & $\frac{2}{n+1} + \frac{n-1}{n+1} \cdot \frac{2}{n}$ & $\frac{2}{n^2+n}$ & $\frac{2n}{2n-1}$ \\
        Conditional Poisson & $\frac{4n-4}{n^2-n+2}$ & $\frac{4n-12}{n^3+n+2}$ & $\frac{n^2-n+2}{n^2-1}$ \\
        \hline
    \end{tabular}
    \caption{An analytical unequal probability example of $n$ elements with $x_1=2$ and $x_i=1$, for $i=2,\dots,n$. The table displays the inclusion probability of the heavy node (node 1) and the deviation among the sampling designs according to the difference and ratio from the str$\pi$ps design.}
    \label{tab:analytical-designs}
\end{table}

The conditional Poisson and draw-by-draw methods are approximations of the str$\pi$ps design and, under certain conditions, the sampling design can have negligible consequences on the application. However, depending on the sensitivity of the application, they might not be suitable to use interchangeably. Specifically, Table~\ref{tab:analytical-designs} shows an analytical example of $n$ elements with sizes $x_1=2$ and $x_i=1, i=2,\dots,n$, i.e. one heavy element with double the size of the other $n-1$ elements. The analysis shows that, in at least one setting, the random sampling designs are different as they lead to different inclusion probabilities for finite $n$. The table also displays the quantified differences among the designs while their asymptotic convergence can also be shown. In particular, the limit of the ratios of both the draw-by-draw design and the conditional Poisson with the str$\pi$ps scheme tends to unity as $n \to \infty$. It is worth mentioning that for single item samples, all these designs are identical.

Finally, from the designs that were mentioned in the beginning of this section, the correspondence can now be stated: interpretation (1) refers to conditional Poisson, interpretation (2) refers to the draw-by-draw method, and interpretation (3) is a str$\pi$ps scheme. All three of these interpretations can be fitted into the description of ``selecting $m$ vertices with probability proportional to their degree'' but all refer to semantically different probabilities that do not coincide with each other.

\section{Variations in the literature and implementations}
\label{sec:related-work}

In this section, we review notable and highly impactful research studies about the BA model and examine the interpretation of the probability proportionality with the node degrees. We also examine the ambiguities that have been previously suggested in the literature and inspect the existing open source implementations of the BA model to uncover the underlying weighted random sampling methodology.

\subsection{Notable BA studies}

The original work of Barabási and Albert formulated the modern version of the \textit{preferential attachment} mechanism, that was previously known as the \textit{Gibrat} principle, the \textit{Yule process}, the \textit{Matthew effect}, or \textit{cumulative advantage} \cite{newman2005power}, and motivated a line of research to expose the underlying properties of this mechanism. However, the absence of explicit definition of the model in respect to the weighted random sampling scheme appears to have led to fragmentation in the literature in terms of the precise interpretation of the model. As a result, the models being studied in the literature have differences from each other as well as from the original definition of the BA model.

For example, \citet{bollobas2001degree} state that
\begin{quote}
    for $m>1$ we add $m$ edges from $u_t$ one at a time \dots,
\end{quote}
which points to a variation of the draw-by-draw selection interpretation (2), as the edges are being added one by one with consecutive draws. This scheme is different than the strict proportionality suggested in the BA model. Furthermore, the phrase
\begin{quote}
    choosing in one go a set $S$ of $m$ earlier vertices
\end{quote}
in the above paper also adds to the confusion as it is unclear whether this refers to generating a sample in one pass (reservoir sampling \cite[Section 3.4.2]{10.5555/270146}) or using a single random draw (whole sampling \cite{10.2307/25048303}), both of which are possible. The authors continue, stating that
\begin{quote}
    it is very natural to allow some of the neighbors to be the same, creating multiple edges in the graph
\end{quote}
which directly contradicts the definition ``\dots~edges that link the new vertex to $m$ \textit{different} vertices already present in the system~\dots''. The same model is also used in a later work of the same authors \citet{bollobas2004diameter}.

In the same paper, the authors identify an ambiguity in the BA model:
\begin{quote}
    it is not clear how the process is supposed to get started,
\end{quote}
referring to the situation of the initial $m_0$ nodes. This ambiguity, however, is not related to the fundamental method of edge selection during the growing process and is asymptotically irrelevant in the BA stationary distribution state \cite{barabasi1999emergence}.

In addition, \citet{batagelj2005efficient} present efficient algorithms for random $\mathcal{G}(n,p)$ and $\mathcal{G}(n,m)$ graphs, small world networks \cite{Watts1998}, and preferential attachment graphs. A connection of the $\mathcal{G}(n,p)$ and $\mathcal{G}(n,m)$ graph generation with random sampling is indirectly given in Section II.A, where the concept of random jumps (see~\cite{10.1145/3147.3165}) is utilized to achieve generation in $\mathcal{O}(n+m)$ time, which is optimal. The random jumps are referred by the authors as ``waiting times''. Although the authors did not identify the connection of preferential attachment with weighted random sampling, they proposed an efficient algorithm for scale-free graphs implementing the BA model. The algorithm works by maintaining a random access data structure $M$, where each vertex resides in $M$ as many times as its degree. One random selection of a single vertex can take place in constant time by querying $M$, and the process yields a multigraph with self-loops, which is also incompatible with the original definition of the BA model.

Another approach towards the analysis of the BA model is given by \citet{hadian2016roll}. The algorithms proposed in that paper are based on the intuition that a lot of nodes have the same degree and, subsequently, the same probabilities of being selected (or included). This situation is more prominent in nodes with lower degrees. A partial sum tree is used to exploit this property, which can be further optimized into a Huffman tree, given that the inclusion probabilities are not uniform. The data structures mentioned in that paper are only applicable when one weighted random selection is desired. The BA algorithm used by the authors generates weighted random vertices sequentially without replacement, a scheme which is equivalent to the draw-by-draw procedure of the interpretation (2) and different than the one indirectly specified in the original BA model. The authors mention that
\begin{quote}
    the computationally intensive part of the algorithm is the degree-proportionate selection, a.k.a. \textit{roulette wheel selection},
\end{quote}
which further demonstrates that the distinction among the weighted random sampling designs appears to be less known in this scientific field as random sampling does not necessarily imply roulette wheel selections.

Finally, in \cite[Chapter 8]{van2016random}, the preferential attachment mechanism is defined in a different way. In particular, for a $m > 1$, each of the $m$ edges are added sequentially and after each addition the degrees of the nodes are updated in the implicit weight data structure before a new selection is made, a concept that is referred to by the author as \textit{intermediate updating of the degrees}. This model can be reduced to a $m=1$ model, where nodes are $m$ consecutive nodes are ``fused'' together to create the final node. Despite the model having a rich-get-richer behavior, it is not exactly equivalent to the original definition of the BA model which disallows self loops and multiple edges.

In conclusion, the literature regarding the analysis of the BA model appears fragmented in respect to its interpretation and, more specifically, the equivalence of the inclusion probabilities of the nodes in the sample of $m$ elements. In particular, the majority of the models are utilizing the interpretation of the draw-by-draw scheme (2) of the selection probabilities with or without replacement, which is both semantically and practically different than the inclusion probability model of the original definition, whether it refers to a multigraph with self loops or not.

\subsection{Open source implementations}

We show that the confusion about the exact interpretation of the probability scheme used in the BA model is also present in its most popular open source implementations. In particular, we observe that the most common open source frameworks with implementations of the BA model use a modification of the algorithm in~\cite{batagelj2005efficient}, that disallows multiple edges by individual selection rejections. This scheme is equivalent to the draw-by-draw scheme without replacement~\cite{yates1953selection} which is semantically different but for large $n$ approximates the str$\pi$ps scheme. We argue that the generated graphs do not exactly abide by the BA model and the mechanism that produces the power law mechanic and, as a result, might not be appropriate for use in some sensitive applications. Despite this, the differences of these \textit{approximation models} with the str$\pi$ps scheme might be negligible for most applications, especially if $n$ is large.

Below we examine the most notable open source implementations of the BA model (the \textit{jGraphT} library, \textit{NetworkX} and \textit{iGraph}) to show their equivalence with the draw-by-draw scheme. We provide a copy of their source code in \ref{sec:source-code} consisting of the code that is being executed when one vertex enters the network that further demonstrates this equivalence.

\paragraph{jGraphT~\cite{michail2019jgrapht}}

Whenever a new edge $v$ -- $u$ is formed, its ends $v$ and $u$ are appended into the \texttt{node} array, which plays the role of a bag, or a multiset; this process is identical to the process of weight updating in the algorithm in~\cite{batagelj2005efficient}. As a result, every vertex in the graph exists inside the \texttt{node} array as many times as its degree and consequently one uniformly random index of the \texttt{node} array corresponds exactly to one weighted random selection. The implementation relies on rejections of duplicate selections in order to achieve a simple graph, via which the equivalence of the draw-by-draw scheme is based on. The memory consumption of the implementation is $\mathcal{O}(nm)$ as the size of the $node$ array increases by $2m$ for each added node and its time complexity is also $\mathcal{O}(nm)$, when not considering the possibility of duplicate elements.

\paragraph{NetworkX~\cite{hagberg2008exploring}}

NetworkX implements the same preferential attachment growing scheme as jGraphT. In particular, the \texttt{\_random\_subset} method implements the weighted selection of the $m$ vertices and operates via the same rejection technique via a set. Similarly, the main method constructs the \texttt{repeated\_nodes} array with nodes inside it as many times as their degree.

\paragraph{iGraph~\cite{csardi2006igraph}}

iGraph implements 3 preferential attachment algorithms labeled as \textit{bag}, \textit{psumtree} and \textit{psumtree-multiple}:
\begin{enumerate}
    \item \textit{bag}: The updating of the weights is done via a bag data structure in the same way as in \textit{jGraphT} and \textit{NetworkX}. The required weighted selection are drawn from the bag in constant time with replacement, resulting in a multigraph.
    \item \textit{psumtree}: This method implements non-linear preferential attachment via a partial sum tree data structure and performs weight updates and random selection in logarithmic time. Using a bag in this case is impossible because of the decimal nature of the weights. It does not allow multiple edges.
    \item \textit{psumtree-multiple}: This method also uses a partial sum tree to generate the graph but allows duplicate edges, resulting in a multigraph too.
\end{enumerate}

\section{First order inclusion probabilities}
\label{sec:first-order}

In this section, we experimentally examine the BA model in the perspective of the first order inclusion probabilities. We implemented three BA generators using three weighted random sampling schemes, each with a different first order probability model, in order to demonstrate their differences in an experimental setting. Moreover, we identify three perspectives under which the properties of the generated graphs are different and show their impact in the BA model. In particular, the divergent properties are the degree distribution (Section~\ref{sec:degree-distribution}), the expectations of individual nodes occupying specific ranks in the degree hierarchy (Section~\ref{sec:rank-probabilities}), and the probabilities of individual vertices overthrowing other vertices during the growing BA model process (Section~\ref{sec:overthrow-probability}). The findings provide evidence that this confusion regarding the first order probabilities has an impact on both a theoretical level and practical applications.

The exact random sampling schemes that we use are the ones mentioned in Section~\ref{sec:random-sampling}:
\begin{enumerate}
    \item Chao's~\cite{chao1982general} algorithm as the str$\pi$ps scheme (abbreviated \textit{$\pi$PS}), which is the first order probability model implied in the original work of Barabási and Albert.
    \item The modification of the draw-by-draw algorithm of~\citet{batagelj2005efficient} (abbreviated DbD) that doesn't produce multigraphs or self-loops and is in use in most open source implementations.
    \item The method directly produced by the definition of conditional Poisson (abbreviated \textit{CP}).
\end{enumerate}

Finally, regarding the initial state of the algorithm, we start the process with a complete graph of $m_0 = m$ nodes. This is important to specify as it addresses the initial clique ambiguities that have been previously addressed.

\subsection{Degree distribution}
\label{sec:degree-distribution}

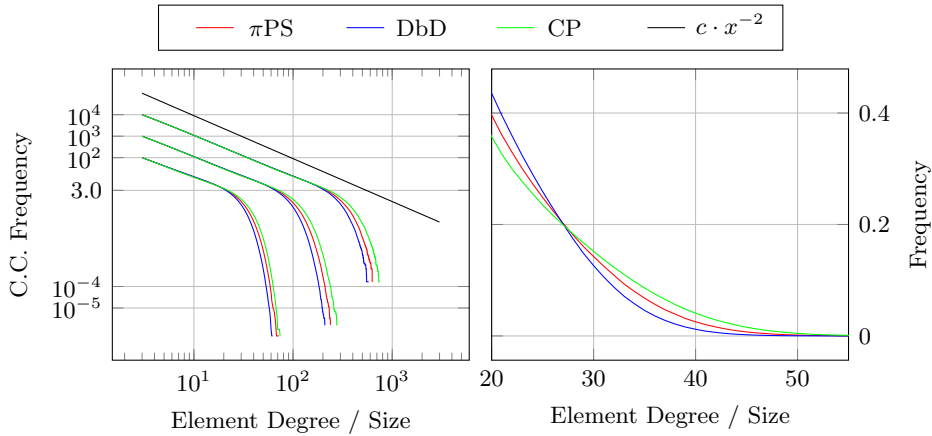
\begin{figure}
    \centering
    \begin{tikzpicture}
        \draw (-0.5,-0.3) -- (7.8,-0.3) -- (7.8,0.3) -- (-0.5,0.3) -- cycle;
        \node at (1, 0) (a) {\footnotesize $\pi$PS};
        \node at (3, 0) (a) {\footnotesize DbD};
        \node at (4.9, 0) (a) {\footnotesize CP};
        \node at (7.1, 0.05) (a) {\footnotesize $c \cdot x^{-2}$};
        \draw [red]   (0,0) -- (0.5,0);
        \draw [blue]  (2,0) -- (2.5,0);
        \draw [green] (4,0) -- (4.5,0);
        \draw [black] (6,0) -- (6.5,0);
    \end{tikzpicture}
    
    \vspace{2mm}
    
    \begin{tikzpicture}[baseline]
        \begin{loglogaxis}[
	        xlabel={Element Degree / Size},
	        ylabel={C.C. Frequency},
	        grid=major,
	        width=0.5\textwidth,
	        ytick={3.0,100,1000,10000,0.00001,0.0001},
	        yticklabels={$3.0$,$10^2$,$10^3$,$10^4$,$10^{-5}$,$10^{-4}$},
	        label style={font=\footnotesize},
	        tick label style={font=\footnotesize},
	        legend style={font=\footnotesize}
        ]
        \addplot [mark=none,red] table [x=degree, y=chao_cd_100, col sep=comma] {figures/data.csv};
        \addplot [mark=none,blue] table [x=degree, y=efraimidis_cd_100, col sep=comma] {figures/data.csv};
        \addplot [mark=none,green] table [x=degree, y=conditional_cd_100, col sep=comma] {figures/data.csv};
        \addplot[black, domain=3:3000, samples=100]{9*10^5*x^(-2)};
        \addplot [mark=none,red] table [x=degree, y=chao_cd_1000, col sep=comma] {figures/data.csv};
        \addplot [mark=none,blue] table [x=degree, y=efraimidis_cd_1000, col sep=comma] {figures/data.csv};
        \addplot [mark=none,green] table [x=degree, y=conditional_cd_1000, col sep=comma] {figures/data.csv};\addplot [mark=none,red] table [x=degree, y=chao_cd_10000, col sep=comma] {figures/data.csv};
        \addplot [mark=none,blue] table [x=degree, y=efraimidis_cd_10000, col sep=comma] {figures/data.csv};
        \addplot [mark=none,green] table [x=degree, y=conditional_cd_10000, col sep=comma] {figures/data.csv};
        \end{loglogaxis}
    \end{tikzpicture}%
    \begin{tikzpicture}[baseline]
        \begin{axis}[
	        xlabel={Element Degree / Size},
	        ylabel={Frequency},
	        grid=major,
	        xmin=20,xmax=55,
	        width=0.5\textwidth,
	        yticklabel pos=right,
	        ylabel near ticks,
	        label style={font=\footnotesize},
	        tick label style={font=\footnotesize},
	        legend style={font=\footnotesize}
        ]
        \addplot [mark=none,red] table [x=degree, y=chao_100, col sep=comma] {figures/data.csv};
        \addplot [mark=none,blue] table [x=degree, y=efraimidis_100, col sep=comma] {figures/data.csv};
        \addplot [mark=none,green] table [x=degree, y=conditional_100, col sep=comma] {figures/data.csv};
        \end{axis}
    \end{tikzpicture}
    \caption{(a) Log-log plot of the complementary cumulative degree distributions of the 3 different random sampling designs. The settings are $n=100,1000,10000$ (indicated by the value of the smallest $x$) and $m=3$. (b) Plot of frequencies of intermediate and large degrees for $n=100$ and $m=3$ in a linear scale.}
    \label{fig:degree-distribution}
\end{figure}

The initial impact of the first order probability confusion is on the degree distribution of the resulting graph. It is known that the expected degree distribution of the BA model follows a power law; an experimental demonstration is shown in Figure~\ref{fig:degree-distribution}a. The figure is a log-log plot of the complementary cumulative degree distributions for $m=3$ and three values for the number of vertices $n$: 100, 1000 and 10000 which are displayed from bottom to top. The plot also displays the reference distribution $x^{-2}$ as the complementary cumulative distribution of the theoretical $x^{-3}$ power law because with a continuum approximation it holds that the cumulative distribution $P(x)$ of $c \cdot x^{-\gamma}$ is
\[
  P(x) = \int_{x}^{\infty} c \cdot x^{-\gamma} = \frac{c}{\gamma-1} x^{-(\gamma-1)} \sim x^{-(\gamma-1)}.
\]
The same plot also displays the results for the 3 different random sampling designs with the tail (the values below 3.5 y-value) being impacted by the very few occurrences.

The differences among the random sampling designs can be better observed in Figure~\ref{fig:degree-distribution}b, which shows in a zoom level the distribution for $n=100$ and $m=3$ in linear axes. The distributions exhibit an interesting behavior regarding the probability mass of the intermediate and higher degrees. In particular, the conditional Poisson design appears to have more vertices with very low or very high degree while the draw-by-draw method results in a distribution with more intermediate degree vertices. Overall, the experiment demonstrates the existence of minor differences on the degree distributions of BA graphs when different random sampling designs are applied on it.

\subsection{Rank probabilities}
\label{sec:rank-probabilities}

Nodes in the BA model graph are usually anonymous, i.e. only the degree distribution is relevant or important. However, there are many graphs that correspond to the same degree distribution and each one has distinct properties. Often, it is desirable to study the graph in the individual vertex level. In this experiment we measure how the degrees are allocated in vertices, and more specifically the degrees of the older nodes, which are usually more important. This scenario would, for example, be applicable when studying the most influential nodes in a social network.

\begin{table}
    \centering
    \begin{tabular}{cc|cc|cc}
    \hline
    \multicolumn{2}{c}{DbD} & \multicolumn{2}{c}{$\pi$PS} & \multicolumn{2}{c}{CP} \\
    \hline
    Node & Top Node Prob. & Node & Top Node Prob. & Node & Top Node Prob. \\
    \hline
    0 & 23.4291\% & 0 & 24.3819\% & 0 & 26.2012\% \\
    1 & 23.4180\% & 1 & 24.3832\% & 1 & 26.1271\% \\
    2 & 23.3936\% & 2 & 24.1091\% & 2 & 26.1779\% \\
    3 & 11.4166\% & 3 & 10.8189\% & 3 & 09.5925\% \\
    4 & 06.3222\% & 4 & 05.8158\% & 4 & 04.6667\% \\
    5 & 03.8164\% & 5 & 03.4175\% & 5 & 02.5734\% \\
    \hline
    \end{tabular}
    \caption{Probabilities of older nodes to occupy the highest degree rank for $n=1000$ and $m=m_0=2$.}
    \label{tab:top-nodes}
\end{table}

According to the BA model, the older nodes are expected to have higher degrees than newly born nodes. We treat the oldest as named nodes and experimentally measure their probabilities to occupy the highest degree rank (the node with the most amount of connections) in respect to the different sampling schemes. We used the settings $n=1000,m=m_0=2$, performed 2,000,000 iterations of the experiment in order to achieve statistical stability, and display the average probabilities in Table~\ref{tab:top-nodes}.

The results provide evidence that the probabilities of the oldest nodes to occupy specific degree ranks is dependent upon the sampling scheme. As a result, in certain cases, for example where the subject of study is the most influential nodes, it is important that there is no ambiguity on the graph generation process. The first 3 nodes have approximately the same probability due to symmetry produced by the initial clique. The table displays only the first 6 nodes to demonstrate the differences of the mot important vertices; the probability drops rapidly as the node ID increases. Profound differences can be observed in these results, as conditional Poisson sampling appears to generate graphs with ``heavier'' old nodes, i.e. nodes that are more likely to remain in the top degree rank after the arrival of newer vertices. Moreover, the same pattern of str$\pi$ps being in-between emerges here as well regarding the ranks of the oldest nodes, while the draw-by-draw design results in the ``lightest'' old nodes.

\subsection{Overthrow probabilities}
\label{sec:overthrow-probability}

Another subtle perspective via which we study the differences among different first order probabilities is targeting the growing component of the process and, more specifically, how the highest degree node changes over time. First, we define a random variable which shows the number of vertices in the growing graph at the point of an \textit{overthrow}. An overthrow indicates a change in the top degree rank and is, more formally, the situation where the vertex with the highest degree is unique (clear heaviest node) and at the exact previous time step there was more than one vertex occupying the highest degree rank (tied heaviest nodes). As a result, an overthrow, as defined here, also includes the situations where a unique top rank of the same node is repeated directly after a tied top rank.

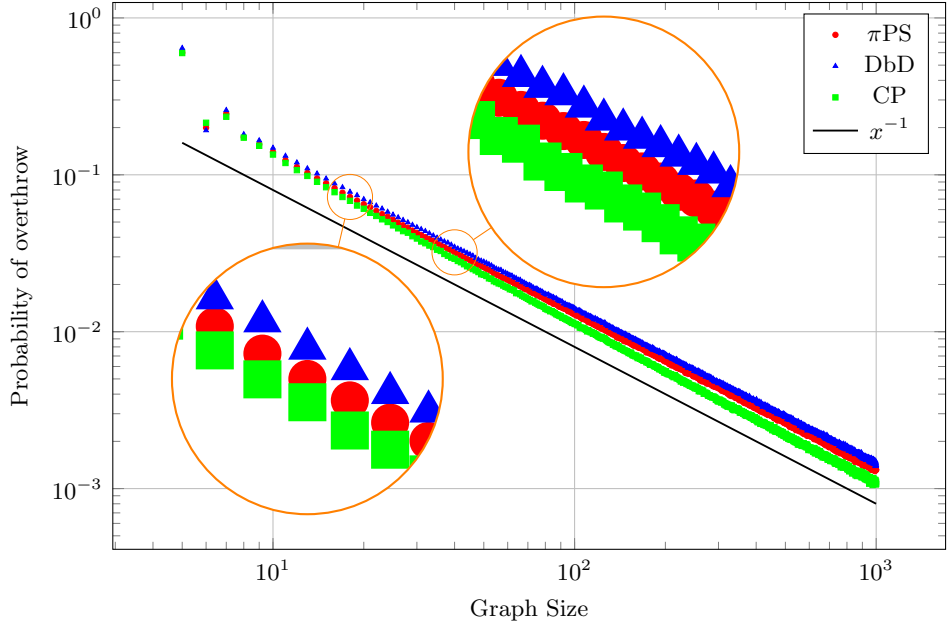
\begin{figure}
    \centering
    \begin{tikzpicture}[
        spy using outlines={circle, magnification=6, connect spies}
    ]
    \begin{loglogaxis}[
            xlabel={Graph Size},
            ylabel={Probability of overthrow},
            grid=major,
            width=\textwidth,
            height=0.7\textwidth,
            label style={font=\footnotesize},
	        tick label style={font=\footnotesize},
	        legend style={font=\footnotesize}
        ]

        \addplot [red,mark=*,only marks,mark size=1] table [x=size, y=chao, col sep=tab] {figures/overthrows.csv};
        \addplot [blue,mark=triangle*,only marks,mark size=1] table [x=size, y=efraimidis, col sep=tab] {figures/overthrows.csv};
        \addplot [green,mark=square*,only marks,mark size=1] table [x=size, y=cp, col sep=tab] {figures/overthrows.csv};

        \addplot[black,domain=5:1000,samples=100,line width=0.25mm]{0.8*x^(-1.00)};

        \coordinate (spypointa) at (axis cs:18,0.072278);
        \coordinate (magnifyglassa) at (axis cs:13,0.005);
        \coordinate (spypointb) at (axis cs:40,0.032038);
        \coordinate (magnifyglassb) at (axis cs:125,0.14);

        \legend{$\pi$PS,DbD,CP,$x^{-1}$}
    \end{loglogaxis}

    \spy [orange, size=3.6cm] on (spypointa) in node[fill=white] at (magnifyglassa);
    \spy [orange, size=3.6cm] on (spypointb) in node[fill=white] at (magnifyglassb);
\end{tikzpicture}
    \caption{Log-log plot of the overthrow probability distribution for $n=1000,m=2$.}
    \label{fig:overthrows}
\end{figure}

It is our expectation that the probability of an overthrow will decline in respect to time. This is because as the node with the highest degree accumulates more edges, the probability of it gaining more edges is further increased and, as a result, the probability of an overthrow is declining over time. Our hypothesis is confirmed by the experiments for $n=1000$, $m=2$ and 2,000,000 repetitions for statistical stability, the results of which are shown in Figure~\ref{fig:overthrows}. The figure shows the probability distribution of an overthrow (y-axis) occurring at specific time points (x-axis) of the BA model growing process. The momentary fluctuations, which are especially noticeable on very small $x$ values are due to the definition of an overthrow that disallows the existence of consecutive overthrows. It is possible that an alternative definition that is not bound by this property would not have this impact on the distribution. Furthermore, the figure also highlights the differences among the random sampling designs with the draw-by-draw design resulting in more ``unstable'' process, i.e. a process with relatively more overthrows. The str$\pi$ps design is again in the intermediate position, which is consistent with the previous findings, and the conditional Poisson generates a BA growing with the least amount of overthrows.

Interestingly, the overthrow probability distribution is also a power law. In fact, the distribution corresponding to the str$\pi$ps design appears remarkably close to the power law distribution $c \cdot x^{-1}$ with a correlation of $\rho=0.99996$ between them. The correlation has been measured for the $x$ values greater or equal than 20 to avoid interference from the initial fluctuations mentioned previously. This observation deserves some attention as, if true, such a complicated process involving the BA model, the weighted random sampling scheme and the overthrow mechanic, produces such a simple and intuitive outcome.

Another interesting property of the approximation -1 of the exponent is that it constitutes a strict boundary on the behavior of another measure, the expected number of overthrows in the asymptotic state, i.e. when $t$ approaches infinity, which is given by $\sum_{t=1}^\infty c \cdot t^{-\gamma}$, where $\gamma$ is the exponent of the overthrow distribution. This expectation converges when $\gamma > 1$ but diverges when $\gamma \le 1$. As a result, the value of the hypothesis that $\gamma = 1$ lies directly on the boundary of asymptotically finite or infinite number of overthrows in a BA graph creation process. We leave the relation of the overthrow distribution with the the power law distribution $x^{-1}$ as an open problem of independent interest; whether they can be analytically linked should be pursued elsewhere.

\section{Higher order inclusion probabilities}
\label{sec:higher-order}

A much more subtle perspective of the behavior of the preferential attachment step of the BA model has to do with the higher order inclusion probabilities of vertices. While the original definition of the BA model indirectly defines the first order inclusion probabilities with the ultimate goal of proving the scale-free behavior of the system, no question is posed about the higher order probabilities, resulting in an ambiguity in the definition. In particular, the $m$-order inclusion probability of the unordered tuple $v_1, v_2, \dots, v_m$ defines the probability that all vertices in the tuple are included in the final selection of vertices during the preferential attachment step.

The first order and second order inclusion probabilities are connected via the formula
\begin{equation}\label{eq:second-to-first}
    \pi_{i} = \sum_{j \in V} \pi_{ij}, \text{ for all } i \text{ in the vertex set } V \text{,}
\end{equation}
where this analogy can be stated equivalently for higher order probabilities. Clearly, given the first order inclusion probabilities, there are in general more than one higher ($m$th) order inclusion probabilities that satisfy the same first order inclusion probabilities, all of which can fit the description of the BA model if not strictly specified. The higher order probability distribution can capture dimensions of the generated graphs that are beyond the degree distribution, for example clustering properties, or path lengths.

In this section, we evaluate this higher order probability ambiguity and show that it has an impact in the resulting graphs. Specifically, we utilize two weighted random sampling schemes of the str$\pi$ps case that differentiate in terms of their second and higher order inclusion probabilities only. These are the ordered systematic sampling~\cite{10.2307/2236209} and the random systematic sampling~\cite{doi:10.1080/01621459.1950.10501130}, both of which belong to the general class of systematic sampling approaches. According to systematic sampling, the items are placed in an implicit weighted array and then $m$ items are selected based on equal weight intervals in that array using a single random variate. The difference between the two sampling designs that we use is that the random systematic performs an implicit shuffle in this array before the selection, while the ordered systematic always assumes the array to have a specific order of the items.

Initially, we demonstrate the differences among these sampling schemes by considering two growing graph generators that are based on the two sampling designs respectively: the ordered systematic generator and the random systematic generator. We show their differences with an analytical example (Section~\ref{sec:generator-probabilities}) for finite $n$ by measuring the probabilities of appearance of any graph that can result from such setting. Moreover, we study these two generators asymptotically (Section~\ref{sec:asymptotic-difference}) and show their divergence with a specific type of graph that is possible in one generator and impossible in the other as a counter example. Finally, we present empirical visualizations (Section~\ref{sec:visualizations}) that illustrate differences among the systematic designs in a perspective that often appears in social network analysis literature.

\subsection{Analytical example of the generator probabilities}
\label{sec:generator-probabilities}

In this analysis, we assume two growing graph generators, each of which uses a different weighted random sampling model for the implementation of the preferential attachment step: the ordered systematic generator and the random systematic generator. For simplicity of the analysis, we consider the order of the vertices of the ordered systematic to be the descending degree order. We analytically show the growing process for $m=2$ and $n=5$, where the outcome of each generator is a list of possible graphs along with their respective probabilities of appearance, and show that for this setting the two particular generators are different as they result in different outcomes. The arguments made in this section are not exclusive for $m=2$ and $n=5$ and it can be shown that generalized analyses are also possible for finite $n$ and $m$.

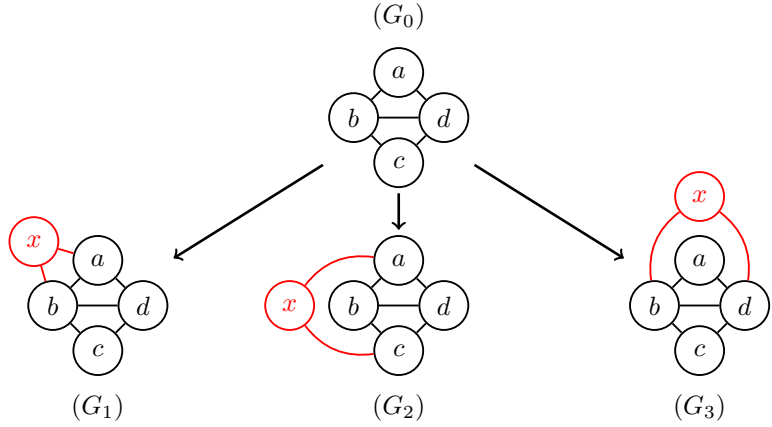
\begin{figure}
    \centering
    \small
    \begin{tikzpicture}
    
    \node[minimum size=2cm] (G0) at (0,-0.5) {}; \node at (0,0.85) {($G_0$)};
    \node[shape=circle,draw=black,line width=0.25mm,minimum size=0.65cm] (G0A) at (0,0.1) {$a$};
    \node[shape=circle,draw=black,line width=0.25mm,minimum size=0.65cm] (G0B) at (-0.6,-0.5) {$b$};
    \node[shape=circle,draw=black,line width=0.25mm,minimum size=0.65cm] (G0C) at (0.6,-0.5) {$d$};
    \node[shape=circle,draw=black,line width=0.25mm,minimum size=0.65cm] (G0D) at (0,-1.1) {$c$};
    \path[-,line width=0.25mm](G0A) edge node[left] {} (G0B);
    \path[-,line width=0.25mm](G0B) edge node[left] {} (G0D);
    \path[-,line width=0.25mm](G0C) edge node[left] {} (G0D);
    \path[-,line width=0.25mm](G0A) edge node[left] {} (G0C);
    \path[-,line width=0.25mm](G0B) edge node[left] {} (G0C);
    
    \node[minimum size=2cm] (G1) at (-4,-3) {}; \node at (-4,-4.35) {($G_1$)};
    \path[->,line width=0.35mm](G0) edge node[left] {} (G1);
    \node[shape=circle,draw=black,line width=0.25mm,minimum size=0.65cm] (G1A) at (-4,-2.4) {$a$};
    \node[shape=circle,draw=black,line width=0.25mm,minimum size=0.65cm] (G1B) at (-4.6,-3) {$b$};
    \node[shape=circle,draw=black,line width=0.25mm,minimum size=0.65cm] (G1C) at (-3.4,-3) {$d$};
    \node[shape=circle,draw=black,line width=0.25mm,minimum size=0.65cm] (G1D) at (-4,-3.6) {$c$};
    \path[-,line width=0.25mm](G1A) edge node[left] {} (G1B);
    \path[-,line width=0.25mm](G1B) edge node[left] {} (G1D);
    \path[-,line width=0.25mm](G1C) edge node[left] {} (G1D);
    \path[-,line width=0.25mm](G1A) edge node[left] {} (G1C);
    \path[-,line width=0.25mm](G1B) edge node[left] {} (G1C);
    \node[shape=circle,draw=red,text=red,line width=0.25mm,minimum size=0.65cm] (G1E) at (-4.85,-2.15) {$x$};
    \path[-,line width=0.25mm,color=red](G1B) edge node[left] {} (G1E);
    \path[-,line width=0.25mm,color=red](G1A) edge node[left] {} (G1E);
    
    \node[minimum size=2cm] (G2) at (0,-3) {}; \node at (0,-4.35) {($G_2$)};
    \path[->,line width=0.35mm](G0) edge node[left] {} (G2);
    \node[shape=circle,draw=black,line width=0.25mm,minimum size=0.65cm] (G2A) at (0,-2.4) {$a$};
    \node[shape=circle,draw=black,line width=0.25mm,minimum size=0.65cm] (G2B) at (-0.6,-3) {$b$};
    \node[shape=circle,draw=black,line width=0.25mm,minimum size=0.65cm] (G2C) at (0.6,-3) {$d$};
    \node[shape=circle,draw=black,line width=0.25mm,minimum size=0.65cm] (G2D) at (0,-3.6) {$c$};
    \path[-,line width=0.25mm](G2A) edge node[left] {} (G2B);
    \path[-,line width=0.25mm](G2B) edge node[left] {} (G2D);
    \path[-,line width=0.25mm](G2C) edge node[left] {} (G2D);
    \path[-,line width=0.25mm](G2A) edge node[left] {} (G2C);
    \path[-,line width=0.25mm](G2B) edge node[left] {} (G2C);
    \node[shape=circle,draw=red,text=red,line width=0.25mm,minimum size=0.65cm] (G2E) at (-1.45,-3) {$x$};
    \path[-,line width=0.25mm,color=red](G2A) edge[bend right] node {} (G2E);
    \path[-,line width=0.25mm,color=red](G2D) edge[bend left] node {} (G2E);
    
    \node[minimum size=2cm] (G3) at (4,-3) {}; \node at (4,-4.35) {($G_3$)};
    \path[->,line width=0.35mm](G0) edge node[left] {} (G3);
    \node[shape=circle,draw=black,line width=0.25mm,minimum size=0.65cm] (G3A) at (4,-2.4) {$a$};
    \node[shape=circle,draw=black,line width=0.25mm,minimum size=0.65cm] (G3B) at (3.4,-3) {$b$};
    \node[shape=circle,draw=black,line width=0.25mm,minimum size=0.65cm] (G3C) at (4.6,-3) {$d$};
    \node[shape=circle,draw=black,line width=0.25mm,minimum size=0.65cm] (G3D) at (4,-3.6) {$c$};
    \path[-,line width=0.25mm](G3A) edge node[left] {} (G3B);
    \path[-,line width=0.25mm](G3B) edge node[left] {} (G3D);
    \path[-,line width=0.25mm](G3C) edge node[left] {} (G3D);
    \path[-,line width=0.25mm](G3A) edge node[left] {} (G3C);
    \path[-,line width=0.25mm](G3B) edge node[left] {} (G3C);
    \node[shape=circle,draw=red,text=red,line width=0.25mm,minimum size=0.65cm] (G3E) at (4,-1.55) {$x$};
    \path[-,line width=0.25mm,color=red](G3C) edge[bend right] node[left] {} (G3E);
    \path[-,line width=0.25mm,color=red](G3B) edge[bend left] node[left] {} (G3E);
    
    \end{tikzpicture}
    \caption{Outcome tree of the preferential attachment growing process for $m=2$. From the initial $G_0$ graph, the 3 outcomes when adding one new node $x$ are shown. $G_1$ appears with probability $\pi_{ab}$ + $\pi_{bc}$ + $\pi_{cd}$ + $\pi_{da}$, $G_2$ with probability $\pi_{ac}$ and $G_3$ with probability $\pi_{bd}$. It is, therefore, evident, that the $m$-order inclusion probabilities are the only measures that determine the generator outcome probabilities.}
    \label{fig:generator-outcomes}
\end{figure}

Initially, as the graph starts with $m=2$ vertices, there is only one outcome when $t=4$ because all 3 outcomes when the forth vertex arrives are isomorphic. This graph is shown as $G_0$ in Figure~\ref{fig:generator-outcomes} which displays the outcome tree of the preferential attachment process. When the fifth vertex arrives, it can be connected to any pair of existing vertices, resulting in any of the graphs $G_1$, $G_2$ or $G_3$, where isomorphic graphs have been merged into $G_1$. The probabilities of appearance of the resulting graphs depend exclusively on the second order inclusion probabilities in this case or the $m$-order inclusion probabilities in the general case. Therefore, for this small scale case, we can analytically calculate the probabilities of appearance of these graphs $P(G_1)$, $P(G_2)$ and $P(G_3)$ for both systematic generators.

\begin{table}
    \centering
    \begin{tabular}{c|c|c|c}
        \hline
         & $P(G_1)$ & $P(G_2)$ & $P(G_3)$ \\
         \hline
         Ordered & $1/5$ & $2/5$ & $2/5$ \\
         Random & $16/30$ & $4/30$ & $10/30$ \\
         \hline
    \end{tabular}
    \caption{Outcome graph probabilities of ordered and random systematic generators.}
    \label{tab:ordered-systematic-comparison}
\end{table}

For the case of ordered systematic and during $t=4$, the sum of the degrees of the vertices is 10 and thus there are 5 equiprobable samples that corresponds to the second order inclusion probabilities $\pi_{12} = 1/5$, $\pi_{13} = 2/5$, $\pi_{14} = 0$, $\pi_{23} = 0$, $\pi_{24} = 2/5$, $\pi_{34} = 0$. As a result, $P(G_1) = 1/5$, $P(G_2) = 2/5$ and $P(G_3) = 2/5$. It is worth mentioning that this result might have been different if the order of the systematic approach was different, for example the age of the nodes instead of their degree. Hence, the order should be part of the sampling scheme in order for the model to be unambiguous. For the case of random systematic, the procedure is equivalent to shuffling the weighted array and performing an ordered systematic selection. The possible shuffles can be then enumerated to determine the second order inclusion probabilities, or these can be calculated using the formulas in~\cite{doi:10.1080/01621459.1966.10480873}. The calculation is omitted in this paper and we present the probabilities of the final graphs which are $P(G_1) = 16/30$, $P(G_2) = 4/30$ and $P(G_3) = 10/30$.

Table~\ref{tab:ordered-systematic-comparison} summarizes these exact probabilities for the two generators and demonstrates the way that the outcomes are differentiated for a small, finite $n$. It is worth mentioning that these two generators are equivalent in terms of their first order inclusion probabilities and equivalent to the str$\pi$ps model that dictates that their inclusion probability is strictly proportional to their degree. As a result, they are both exactly compatible with the definition of the BA model and the preferential attachment mechanism.

\subsection{Asymptotic impact of higher order probabilities}
\label{sec:asymptotic-difference}

The asymptotics of the BA model have been extensively studied in the literature that examine the model in a stationary process perspective. It is known that a property that can have a significant impact for small $n$ might asymptotically have no effect. One such property is the initial graph from which the process starts, that is asymptotically irrelevant. A natural question is whether this situation occurs for the different higher order probability models as well.

\begin{figure}
    \centering
    \begin{tikzpicture}
        \draw[black,line width=0.25mm] (60:2) arc (60:360:20mm);
        \node[fill=white] (center) at (0,0) {$v_0$};
        \foreach \phi in {1,...,6}{
        \node[fill=white] (v_\phi) at (360/6 * \phi:2cm) {$v_\phi$};
        \draw[black,line width=0.25mm] (v_\phi) -- (center);
        }
    \end{tikzpicture}
    \caption{A wheel graph that is being produced by the random systematic generator that is missing one of its arc edges. This situation is possible for $m=2$ by always selecting the center node and the newest node as connections for newborn nodes. For example, the next vertex $v_7$ in this graph will be connected with $v_0$ and $v_6$. This situation, although extremely unlikely, is possible to occur in the random systematic generator, no matter how large $n$ gets.}
    \label{fig:wheel-graph}
\end{figure}
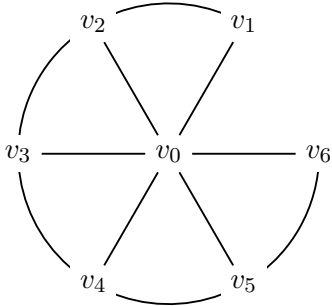

In this section, we prove that the higher order inclusion probabilities can have an asymptotic impact on the graph generator by presenting a counter example on a specific setting. In particular, we utilize the ordered systematic and random systematic generators but in this analysis we consider the order of the ordered generator to be the age of the nodes, where the newest node is attached to the right side of the implicit weighted array of the algorithm. The graph used as a counter example is the graph illustrated in Figure~\ref{fig:wheel-graph}, which we prove that is possible to occur in the random systematic generator but impossible in the ordered systematic generator for all $n$ that are greater than a specific finite value.

\begin{figure}
    \centering
    \begin{tikzpicture}
        \draw (0,0) rectangle (1,1) node[pos=.5] {$\dots$};
        \draw (1,0) rectangle (2,1) node[pos=.5] {$t-1$};
        \draw (2,0) rectangle (3,1) node[pos=.5] {$3$};
        \draw (3,0) rectangle (4,1) node[pos=.5] {$3$};
        \draw (4,0) rectangle (5,1) node[pos=.5] {$3$};
        \draw (5,0) rectangle (6,1) node[pos=.5] {$\dots$};
        \draw (6,0) rectangle (7,1) node[pos=.5] {$3$};
        \draw (7,0) rectangle (8,1) node[pos=.5] {$3$};
        \draw (8,0) rectangle (9,1) node[pos=.5] {$3$};
        \draw (9,0) rectangle (10,1) node[pos=.5] {$3$};
        \draw (10,0) rectangle (11,1) node[pos=.5] {$2$};
        
        \draw (0,1.1) rectangle (1,2.1) node[pos=.5] {$\dots$};
        \draw (1,1.1) rectangle (2,2.1) node[pos=.5] {$t-1$};
        \draw (2,1.1) rectangle (3,2.1) node[pos=.5] {$3$};
        \draw (3,1.1) rectangle (4,2.1) node[pos=.5] {$3$};
        \draw (4,1.1) rectangle (5,2.1) node[pos=.5] {$3$};
        \draw (5,1.1) rectangle (6,2.1) node[pos=.5] {$\dots$};
        \draw (6,1.1) rectangle (7,2.1) node[pos=.5] {$3$};
        \draw (7,1.1) rectangle (8,2.1) node[pos=.5] {$3$};
        \draw (8,1.1) rectangle (9,2.1) node[pos=.5] {$3$};
        \draw (9,1.1) rectangle (10,2.1) node[pos=.5] {$2$};
    \end{tikzpicture}
    \caption{Ordered systematic sampling weights for the wheel graph of Figure~\ref{fig:wheel-graph}. A newborn node has to connect with the center node of degree $t-1$ and the node with degree 2. As a result, the sum of the degrees between the center and the node with degree 2 increases by 3 each time a newborn node is added to the network.}
    \label{fig:ordered-systematic-wheel}
\end{figure}

We consider the state of the ordered systematic generator to be an array of degrees where the value of some index $i$ shows the degree of the node that entered at time $i$. It is easy to show that there is only one vertex with degree $t-1$ (the center), two vertices with degree 2, all other vertices have degree 3, and that the last element in the array always has degree 2 because it always enters last. As a result, after the addition of a newborn node that connects with the center and the node that was previously last, the sum of the weights between the center node and the node with degree 2 is increased by 3. This process is illustrated in Figure~\ref{fig:ordered-systematic-wheel}. In contrast, the skip of the ordered systematic process increases by half the added weight, which is 2. Therefore, it is clear that as the required distance increases faster than the systematic skip, there exists some finite $t$ after which the second order inclusion probability of the center and the last node is zero. As a result, the graph illustrated in Figure~\ref{fig:wheel-graph} is impossible to occur via the ordered systematic generator for some values of $n$ greater than a constant. It is easy to see that even if the center is allowed to be selected together with any of the two nodes of degree 2, the average weight sum between the center and the first node of degree 2 still increases by 3.

Conversely, the situation of the selection of the center node with the last entered node together is always possible in the random systematic generator, as it is equivalent to randomizing the order at which nodes appear in the weight array and, hence, there exists no finite $n$ at which the second order inclusion probability of the center with the last entered node becomes zero. A more general finding can be formulated based on the observation that on ordered systematic generation there exists constant $t_0$ after which the illustrated wheel graph can no longer appear. As a result, it also follows that all graphs that can be generated from a wheel graph of $t_0$ vertices, which may not be wheel graphs themselves (for example a new vertex $v_7$ connecting with $v_3$ and $v_5$), are also impossible to occur, despite the fact that there could be isomorphic graphs in other paths of the generation tree.

The analysis in this section shows that two generators with random sampling schemes being equivalent in terms of their first order inclusion probabilities might not be equivalent overall, even asymptotically. As a result, the BA model cannot be defined based exclusively on the first order inclusion probabilities, as was done in the original definition, but needs to be refined to account for the higher order probabilities.

\subsection{Empirical visualizations}
\label{sec:visualizations}
 
In this section, we perform an empirical analysis and attempt to imprint the differences among the systematic designs in a visually appealing way. In particular, we present force directed visualizations of the ordered and random systematic designs that highlights the clustering features of each generator and provides indications of the pairwise similarities among the nodes.

We utilize the random systematic and the ordered systematic design where the vertices assume an age ordering, but we also enrich this section with a special kind of ordered systematic sampling approach. According to this new approach that refers to $m=2$, newborn nodes are inserted into the implicit weighted array in 4 deterministic positions in succession that repeat after 4 nodes have been added:
\begin{enumerate}
    \item [1.] the left side of the array
    \item [2.] the right side of the array
    \item [3 - 4.] the middle of the array (twice)
\end{enumerate}
Here, the middle of the array is the index for which the sum of weights of all elements to the left is equal to the sum of weights of all elements to the right or the index which optimizes the difference between these quantities if there is no exact middle. The intuition behind this new design is that a pair of nodes that reside in the array with distance equal to half the sum of all weights will approximately maintain this property after 4 newborn nodes have been added. As a result, these pairs of vertices that typically gain edges together when newborn nodes enter the network, will continue to exhibit this behavior asymptotically. In contrast, other pairs of vertices whose distance is much lower or higher than half the degree sum, will likely never gain common neighbors in the future. Hence, we call this approach the \textit{romantic systematic generator}. The romantic systematic generator, as well as the ordered and random systematic, are str$\pi$ps generators with different higher order probability profiles.

The analysis we perform in this section is motivated by an inherent empirical impact of the higher order inclusions probabilities in the preferential attachment generator. In particular, we are looking for pairs of vertices that gain edges together after a newborn node has been added, as opposed to individual vertices gaining edges. A graph transformation that is able to imprint this property exists in a social network analysis context and is typically referred to as \textit{projection}; projection methods have been used before in real social networks with significant success~\cite{STAMATELATOS2020102172}. In the analysis of this section we utilize the Jaccard projection, which transforms the original graph into a complete weighted graph of equal order, where the weight of an edge between node $i$ and $j$ is equal to the Jaccard index of their respective neighbor sets: $w_{ij} = (a_i \cap a_j) / (a_i \cup a_j)$. This method is able to imprint the pairwise relations among the nodes and uncover a variety of subtle properties that are more difficult to obtain given only the (first order or direct) edges among the vertices.

The concept of projections is also related to the idea of \textit{proximity}~\cite[Section 2.1]{8294302} that, similar to inclusion probabilities, comprises first order proximity and higher order proximity. The first order proximity shows direct relations (edges) among pairs of vertices while the second order shows the relation in respect to their direct neighborhoods. The Jaccard index belongs to the second order proximity measures and perfectly fits the concept of pairwise node relations that we attempt to demonstrate.

\begin{figure}
    \centering
    \includegraphics[width=0.3\textwidth]{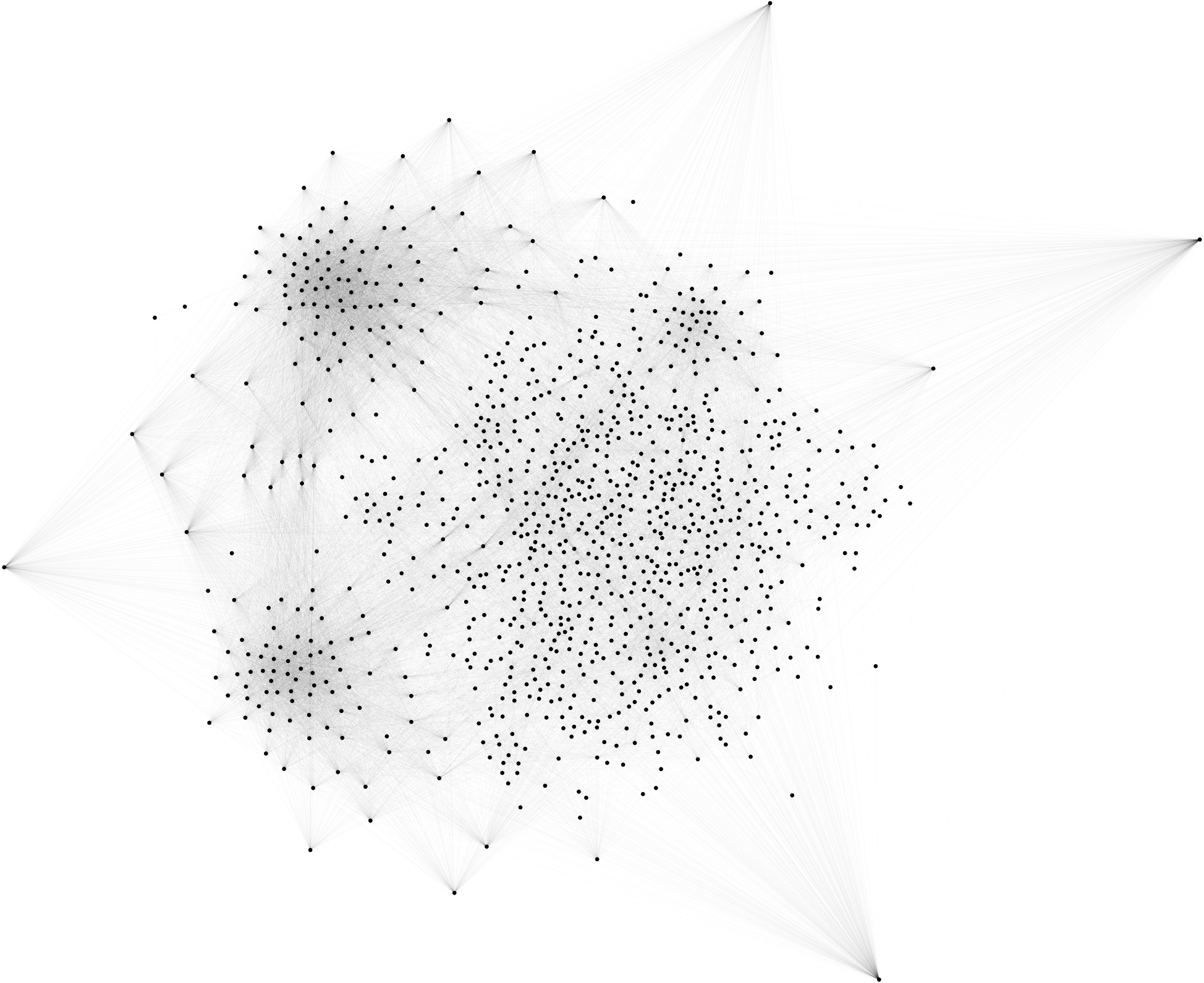}
    \includegraphics[width=0.3\textwidth]{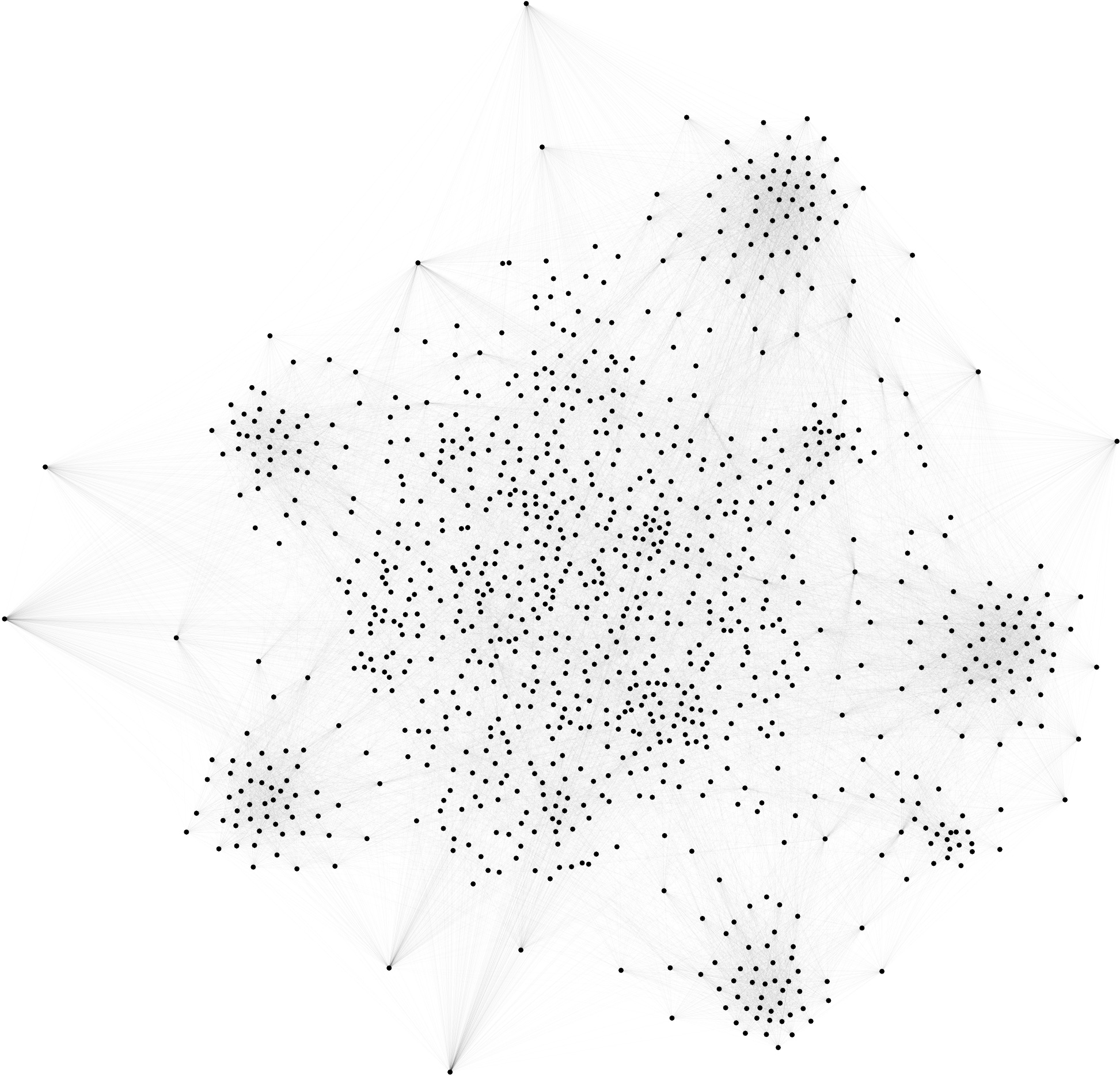}
    \includegraphics[width=0.3\textwidth]{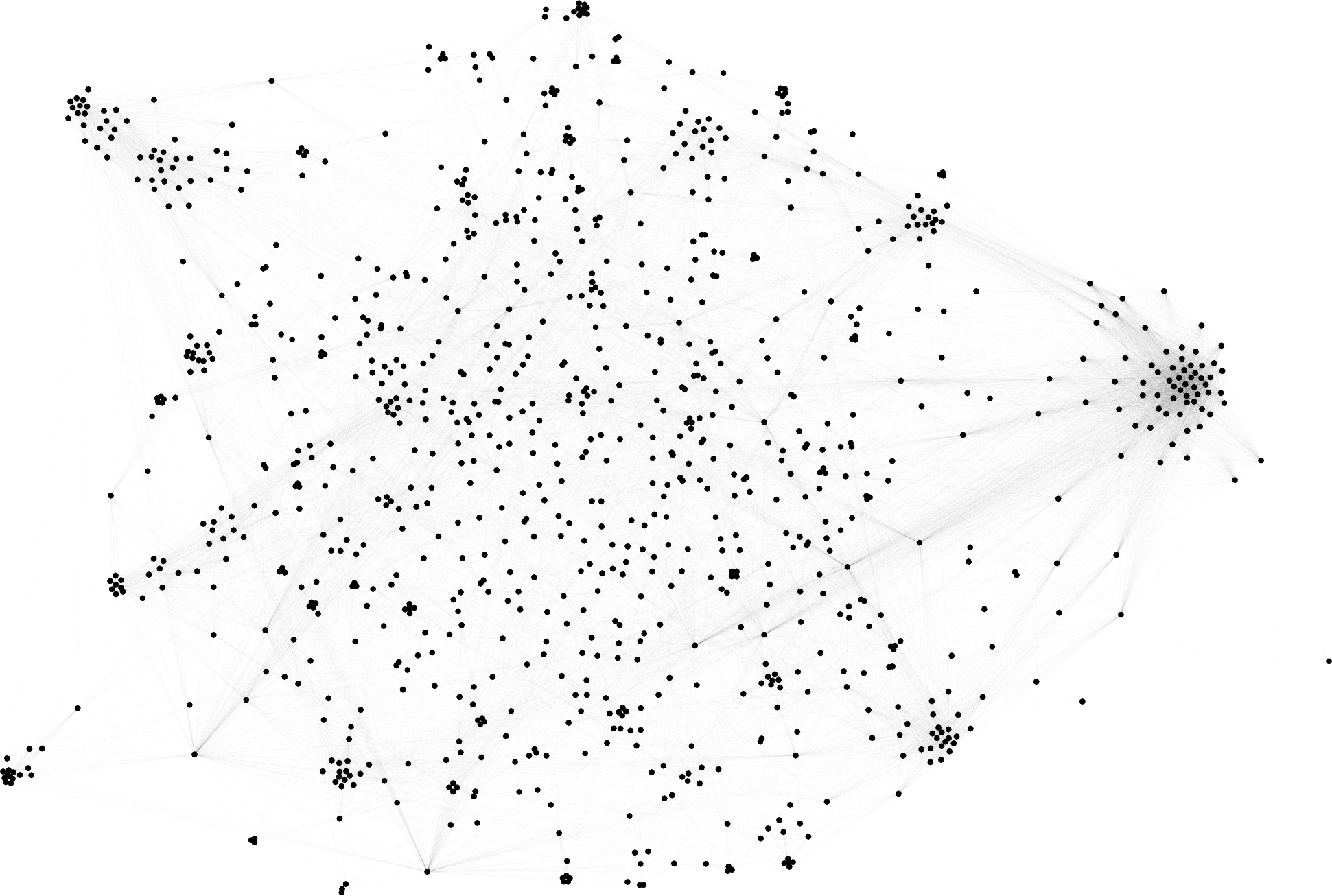}
    \caption{Force directed visualizations of the Jaccard projections of one random systematic graph (left), one ordered systematic (middle) and one romantic systematic (right).}
    \label{fig:visualizations}
\end{figure}

Figure~\ref{fig:visualizations} shows force directed visualizations of the three designs involved in this analysis. We selected one random graph from each design with $n=1000$ and $m=2$ and applied the Jaccard projection before using the final output into the visualizer. The force directed visualization tool optimizes the placement of the vertices in such a way that nodes with high Jaccard similarity are generally closer in the layout. The layouts reveal interesting properties regarding the pairwise relation of vertices and the clustering of the networks, in comparison with each other.

The random systematic graph contains minimal clustering features with very few, dense clusters. In contrast, the ordered systematic projection appears to contain more and smaller clusters which is due to the possible pairs of nodes capable of gaining edges together having a high probability of maintaining this state after a new node enter the network. This is made possible because, despite the possible pairs sliding as new nodes enter to the right of the array, this change is gradual and depending on the degree of the nodes involved, the respective second order inclusion probability could remain unaffected. Finally, the romantic systematic projection has a unique behavior and illustrates why we synthesized its operation. The clusters of this layout are visibly much smaller than the other two projections and the vertices appear overall more evenly geometrically distributed. This result is not surprising given that the method was designed for a vertex to maintain its preferences regarding the selection pairs over the graph creation duration.
As a result, most of the vertex pairs have no common neighbors at all while the probability of pairs of nodes that already have common neighbors to gain new ones is very high. The visible clusters in this layout can be explained by the high degree nodes which, due to their size in the array could have multiple pairing nodes.

In conclusion, with this analysis we empirically demonstrate the differences of three preferential attachment generator designs in respect to the pairwise similarities among the nodes via the visualization of their Jaccard projections. Despite these designs being all strictly $\pi$ps, the differences in the visualizations are profound and are attributed to the differences in the higher order probability profile of each generator.

\section{Concise definition of the BA model}
\label{sec:concise-definition}

Typically, the BA model is defined via the parameters $n$, $m$, $m_0$ and less commonly, in non-linear models, the $a$ parameter. As we have shown in Section~\ref{sec:higher-order}, this definition is not sufficient to adequately describe the process and is open to interpretation. In particular, we have seen that the higher order probabilities create an ambiguity which needs to be addressed. As a result, in this section, we suggest a concise definition of the BA model in random sampling terminology based on the findings of this paper.

Initially, regarding the confusion about the proportionality of the first order inclusion probabilities with the degrees of the vertices, it can be easily deduced from the original definition of the BA model that it refers to a str$\pi$ps model, where the inclusion probabilities of the nodes are exactly proportional to their degrees. The sampling design is without replacement and with constant sample size in order to abide by the constant $m$ parameter and result in a simple graph, i.e. a graph without loops or multiple edges. Moreover, the $m$-order inclusion probabilities need to be specified in order for the relevant ambiguity to be resolved and the model to be reproducible. Once two weighted random sampling methods are identical in terms of their $m$-order inclusion probabilities, they are identical in terms of the BA model with $m$ number of edges added in each step. It is worth mentioning that the $m$-order inclusion probabilities define any other higher order probability lower than $m$ via the generalization of the recursive Equation~\ref{eq:second-to-first}, that holds for fixed sample size. Hence, only the $m$-order inclusion probabilities need to be specified.

The concept of equivalence in weighted random sampling designs is highlighted by \citet{brewer_sampling_1983}:
\begin{quote}
    Two procedures belong to the same equivalence class when the joint probabilities of inclusion of all possible combinations of units are identical.
\end{quote}
Known classifications of weighted random sampling schemes in equivalence classes are also presented in that survey. Based on this definition, two sampling methods in the same equivalence class will produce identical growing preferential attachment graph generators. The reverse might not necessarily be true, for example when they are equivalent only in terms of their $m$-order probabilities (but not higher). Finally, we believe that in most settings, unless some special condition apply, a plausible assumption would be for all $m$-order inclusion probabilities to be strictly positive.

Despite the BA model becoming unambiguous by specifying the $m$-order inclusion probabilities, there is a need to emphasize that sometimes this might be challenging. For certain weighted random sampling methods, the higher order inclusion probabilities are not known or there is no known formula to compute them. In such cases, the ambiguity of the model is resolved by describing the exact sampling algorithm used in the preferential attachment step. We finally note that the same arguments hold for the non-str$\pi$ps case. Despite not being exactly compatible with the original definition of the BA model, the ambiguity still exists without the specification of the $m$-order inclusion probabilities, although the first order probabilities might not be exactly proportional to the degree.

\section{Conclusions}

In this paper we examined the BA model in the new perspective of inclusion probabilities by studying the preferential attachment mechanism with the field on unequal probability random sampling. In particular, we studied separately two cases of inclusion probabilities: the first order and the higher order inclusion probabilities. We have discovered that for the first order probability case, both the literature and the respective open source implementations of the model are fragmented. Despite the original work from Barabási and Albert indirectly points to a str$\pi$ps model, relevant models appear to interpret the inclusion probabilities of vertices in a way that is only an approximation of the exact model. On the higher order case, we discovered that there is a subtle ambiguity regarding the probabilities that any tuple of vertices will be selected during the preferential attachment step, which results in the model being open to interpretation. We showed that this ambiguity can have an impact on the properties of the resulting BA graph, even asymptotically. To this extent, we proposed a new and concise definition of the model that addresses this ambiguity.

In an abstract level, via this work, we suggest that the BA model is in fact not a single model with particular properties but a family of models, where each one might have different properties. Specifically, we argue that a mention in a model in this family must be accompanied by the $m$-order inclusion probabilities, or the exact scheme, or the equivalence class of the random sampling method used in the preferential attachment step. Additional novelties were presented during our analyses that could be of independent interest, for example the concept of overthrow probabilities that appears to create a power law distribution.

Future work should pursue insights into both cases of inclusion probabilities. For the former case, efficient algorithms which satisfy the str$\pi$ps property should be looked into, as current literature is based on draw-by-draw selection models and not the exact case. Unexpected implications of existing implementations on practical applications should also be assessed, even if their impact is negligible. For the case of higher order probabilities, a complete analysis of the impact of different models should be pursued. Several properties have been studied for individual models and should be examined in relation to the concept of higher order inclusion probabilities.

As a closing point we would like to remark that other applications of unequal probability random sampling might also be affected by the phenomena we discuss in this paper. While we showed the close association between weighted random sampling and the preferential attachment mechanism, our analysis might be relevant to potentially any application involving a random selection of more than 1 item from a population based on given weights, including fields outside of computer sciences.

\section*{Acknowledgements}
This work has been co-financed by the European Union and Greek national funds through the Operational Program Competitiveness, Entrepreneurship and Innovation, under the call RESEARCH -- CREATE -- INNOVATE (project code: T1EDK-02474, grant no.: MIS 5030446).

\bibliographystyle{apalike}
\bibliography{main}

\appendix

\section{Source Code of Open Source Libraries}
\label{sec:source-code}

\subsection{jGraphT (Java, Version 1.5.0)}
\label{sec:jgrapht-code}

\begin{minted}[frame=lines,framesep=2mm,baselinestretch=1.2,fontsize=\footnotesize]{java}
List<V> newEndpoints = new ArrayList<>();
int added = 0;
while (added < m) {
    V u = nodes.get(rng.nextInt(nodes.size()));
    if (!target.containsEdge(v, u)) {
        target.addEdge(v, u);
        added++;
        newEndpoints.add(v);
        if (i > 1) {
            newEndpoints.add(u);
        }
    }
}
nodes.addAll(newEndpoints);
\end{minted}

\subsection{NetworkX (Python, Version 2.4)}
\label{sec:networkx-code}

\begin{minted}[frame=lines,framesep=2mm,baselinestretch=1.2,fontsize=\footnotesize]{python}
def main():
    while source < n:
        # Add edges to m nodes from the source.
        G.add_edges_from(zip([source] * m, targets))
        # Add one node to the list for each new edge just created.
        repeated_nodes.extend(targets)
        # And the new node "source" has m edges to add to the list.
        repeated_nodes.extend([source] * m)
        # Now choose m unique nodes from the existing nodes
        # Pick uniformly from repeated_nodes (preferential attachment)
        targets = _random_subset(repeated_nodes, m, seed)
        source += 1
    return G

def _random_subset(seq, m, rng):
    targets = set()
    while len(targets) < m:
        x = rng.choice(seq)
        targets.add(x)
    return targets
\end{minted}

\subsection{iGraph (C, Version 0.8.2)}
\label{sec:igraph-code}

\subsubsection{Method bag}

\begin{minted}[frame=lines,framesep=2mm,baselinestretch=1.2,fontsize=\footnotesize]{c}
for (j = 0; j < no_of_neighbors; j++) {
    long int to = bag[RNG_INTEGER(0, bagp - 1)];
    VECTOR(edges)[resp++] = i;
    VECTOR(edges)[resp++] = to;
}    
\end{minted}

\subsubsection{Method psumtree}

\begin{minted}[frame=lines,framesep=2mm,baselinestretch=1.2,fontsize=\footnotesize]{c}
/* Select nodes */
for (j = 0; j < no_of_neighbors; j++) {
    sum = igraph_psumtree_sum(&sumtree);
    igraph_psumtree_search(&sumtree, &to, RNG_UNIF(0, sum));
    VECTOR(degree)[to]++;
    IGRAPH_CHECK(igraph_vector_push_back(&edges, i));
    IGRAPH_CHECK(igraph_vector_push_back(&edges, to));
    edgeptr += 2;
    IGRAPH_CHECK(igraph_psumtree_update(&sumtree, to, 0.0));
}
/* Update probabilities */
for (j = 0; j < no_of_neighbors; j++) {
    long int nn = (long int) VECTOR(edges)[edgeptr - 2 * j - 1];
    IGRAPH_CHECK(igraph_psumtree_update(&sumtree, nn,
        pow(VECTOR(degree)[nn], power) + A));
}
\end{minted}

\subsubsection{Method psumtree-multiple}

\begin{minted}[frame=lines,framesep=2mm,baselinestretch=1.2,fontsize=\footnotesize]{c}
/* Select nodes */
for (j = 0; j < no_of_neighbors; j++) {
    igraph_psumtree_search(&sumtree, &to, RNG_UNIF(0, sum));
    VECTOR(degree)[to]++;
    VECTOR(edges)[edgeptr++] = i;
    VECTOR(edges)[edgeptr++] = to;
}
/* Update probabilities */
for (j = 0; j < no_of_neighbors; j++) {
    long int nn = (long int) VECTOR(edges)[edgeptr - 2 * j - 1];
    IGRAPH_CHECK(igraph_psumtree_update(&sumtree, nn,
        pow(VECTOR(degree)[nn], power) + A));
}
\end{minted}

\end{document}